\documentclass[a4paper,fleqn]{cas-dc}
\usepackage[authoryear,longnamesfirst]{natbib}
\usepackage{enumitem}

\def\tsc#1{\csdef{#1}{\textsc{\lowercase{#1}}\xspace}}
\tsc{WGM}
\tsc{QE}
\tsc{EP}
\tsc{PMS}
\tsc{BEC}
\tsc{DE}
\usepackage{graphicx}
\usepackage{caption}
\usepackage{array}
\usepackage{times}
\begin{document}

\let\printorcid\relax
\let\WriteBookmarks\relax
\def\floatpagepagefraction{1}
\def\textpagefraction{.001}
% \shorttitle{Leveraging social media news}
% \shortauthors{J.K. Krishnan et~al.}
\setlength{\abovedisplayskip}{5pt} % 公式上方间距
\setlength{\belowdisplayskip}{5pt} % 公式下方间距
\title [mode = title]{Automated and Scalable SEM Image Analysis of Perovskite Solar Cell Materials via a Deep Segmentation Framework}

\author[1]{Jian Guo Pan}

\author[1]{Lin Wang}
\ead{1000484365@smail.shnu.edu.cn}

\author[2]{Xia Cai \corref{cor1}}  % 通信作者

\address[1]{Shanghai Normal University, 100 Guilin Road, Shanghai 200234, China}
\address[2]{Fudan University, 220 Handan Road, Shanghai 200433, China}

\cortext[cor1]{Corresponding author}
\begin{abstract}
%optimizing material properties, controlling film quality, and improving solar cell efficiency
%is particularly effective for capturing fine-grained crystal structures, complex boundaries, and multi-scale features in challenging scenarios.
Scanning Electron Microscopy (SEM) is indispensable for characterizing the microstructure of thin films during perovskite solar cell fabrication. Accurate identification and quantification of lead iodide and perovskite phases are critical because residual lead iodide strongly influences crystallization pathways and defect formation, while the morphology of perovskite grains governs carrier transport and device stability. Yet current SEM image analysis is still largely manual, limiting throughput and consistency. Here, we present an automated deep learning-based framework for SEM image segmentation that enables precise and efficient identification of lead iodide, perovskite and defect domains across diverse morphologies. Built upon an improved YOLOv8x architecture, our model named PerovSegNet incorporates two novel modules: (i) Adaptive Shuffle Dilated Convolution Block, which enhances multi-scale and fine-grained feature extraction through group convolutions and channel mixing; and (ii) Separable Adaptive Downsampling module, which jointly preserves fine-scale textures and large-scale structures for more robust boundary recognition. Trained on an augmented dataset of 10,994 SEM images, PerovSegNet achieves a mean Average Precision of 87.25\% with 265.4 Giga Floating Point Operations, outperforming the baseline YOLOv8x-seg by 4.08\%, while reducing model size and computational load by 24.43\% and 25.22\%, respectively. Beyond segmentation, the framework provides quantitative grain-level metrics, such as lead iodide/perovskite area and count, which can serve as reliable indicators of crystallization efficiency and microstructural quality. These capabilities establish PerovSegNet as a scalable tool for real-time process monitoring and data-driven optimization of perovskite thin-film fabrication.The source code is available at: \href{https://github.com/wlyyj/PerovSegNet/tree/master}{https://github.com/wlyyj/PerovSegNet/tree/master}.
\end{abstract}

\begin{keywords}
Perovskite solar cells \sep Scanning Electron Microscopy (SEM) \sep Deep Learning Segmentation \sep YOLOv8-based Framework \sep Grain Boundary Analysis \sep Lead Iodide and Perovskite Quantification
\end{keywords}

\maketitle
\section{Introduction}
\label{sec1}
Perovskite solar cells (PSCs) have progressed rapidly since their initial efficiency of 3.9\% in 2009, with the state-of-the-art power conversion efficiency (PCE) now exceeding 27\%~\citep{NREL2025}. Their remarkable efficiency gains, coupled with low fabrication cost and solution processability, underscore their potential for scalable and sustainable energy generation~\citep{sun2022high}. A key factor affecting the performance and long-term stability of PSCs is the microscopic morphology of the perovskite thin film~\citep{xie2020revealing,zeng2017morphology}. These films are typically polycrystalline, with crystallinity, grain size and boundary characteristics playing critical roles in charge transport and recombination~\citep{yang2023recent}. Uniform grains and well-defined boundaries enhance carrier mobility, while defects and irregular grain interfaces serve as trap sites that degrade performance and accelerate material degradation under environmental stressors~\citep{urbina2020balance}. Therefore, precise characterization and quantification of film morphology, including grain boundaries, defects, and secondary phases like residual lead(II) iodide, which typically arises in iodine-lead-based perovskite systems due to incomplete precursor conversion, are essential for optimizing the fabrication process and improving both efficiency and stability. Scanning electron microscopy (SEM) is widely employed for high-resolution morphological analysis of perovskite films~\citep{rothmann2017microstructural,hidalgo2019imaging,mishra2025microscopic}, which can provide nanoscale insight into surface structure, enabling the identification of key microstructural features. However, manual interpretation of SEM data is labor-intensive, prone to subjective bias, and unsuitable for high-throughput analysis~\citep{tuo5318317intelligent,kuznetsova2024expanding}. As a result, automated and accurate image segmentation techniques have become essential tools for materials characterization~\citep{stier2024materials}.

Recent advances in image segmentation methods fall into two main categories: traditional machine learning and deep learning (DL)~\citep{minaee2021image}. Traditional approaches, such as support vector machines~\citep{jakkula2006tutorial} and random forests~\citep{rigatti2017random}, rely heavily on hand-crafted features and thresholding rules. For instance, Yang et al. introduced a progressive grayscale thresholding method to analyze particle morphology in perovskite films, achieving reasonable segmentation results~\citep{yang2024grain}. However, such methods struggle with blurry boundaries and low-contrast regions, often misclassifying lead iodide clusters or merging adjacent grains~\citep{de2024morphological}. In contrast, DL methods, especially convolutional neural networks~\citep{chauhan2018convolutional} and YOLO-based models~\citep{jiang2022review}, offer significant advantages for analyzing complex microstructures in materials science, which can automatically learn hierarchical representation and perform efficient inference to enable precise and scalable segmentation. For instance, Lilliu et al. combined nano-focused X-ray diffraction with SEM imaging and utilized DL to identify grain boundaries~\citep{lilliu2016mapping}. While this significantly enhanced segmentation accuracy, the method lacked scalability when applied to large datasets involving perovskite and lead iodide. Similarly, Jin et al. employed Electron Backscatter Diffraction alongside SEM imaging to assess grain size~\citep{jin2022avoid}. However, the method relied heavily on manual parameter adjustments, compromising its generalizability across diverse perovskite morphologies. In another study, Hidalgo et al. incorporated 3D reconstruction with DL to improve segmentation precision~\citep{hidalgo2019imaging}. Despite its accuracy, the approach was computationally inefficient when processing complex, multilayered structures, rendering it impractical for real-time or high-throughput applications. YOLOv8, a state-of-the-art real-time detection model, integrates transformer-based modules and an anchor-free design for robust object detection. Despite its strengths, YOLOv8 struggles with the accurate delineation of fine-grained features~\citep{vijayakumar2024yolo}, such as grain boundaries and small defect regions, and its performance remains highly dependent on the availability of large, high-quality annotated datasets~\citep{ali2024yolo,ma2025addressing}. This dependency poses a significant challenge in the context of SEM imaging, where images often suffer from low contrast, background noise, and ambiguous structural boundaries~\citep{rothmann2017microstructural,yao2023low,rothmann2020atomic}, making manual annotation both labor-intensive and error-prone.

To address these challenges, we introduce PerovSegNet, a lightweight and high-precision segmentation framework tailored for SEM image analysis of perovskite and lead iodide materials. Built upon the YOLOv8-seg backbone, PerovSegNet incorporates two novel modules: Adaptive Split-Dilated Convolution with Shuffle Block (ASDCB), for enhancing feature fusion via multi-branch dilated convolutions and attention-based channel recalibration, and Separable Adaptive Downsampling module (SAD), for capturing both fine-scale (high-frequency) textures and large-scale (low-frequency) structures. To overcome the scarcity of annotated SEM data, we construct an augmented PerovData dataset of 10,994 images from 40 high-quality manually labeled samples and employ Uniform Manifold Approximation and Projection (UMAP)-based analysis to verify feature separability and morphological diversity. Beyond achieving state-of-the-art segmentation accuracy and efficiency relative to baseline methods, PerovSegNet enables automated quantification of grain area and count via post-segmentation analysis, thereby offering a robust tool for assessing PCE and supporting materials optimization during device fabrication.

\section{Method}
\begin{figure*}[htbp]
\centering
\includegraphics[width=1.0\textwidth]{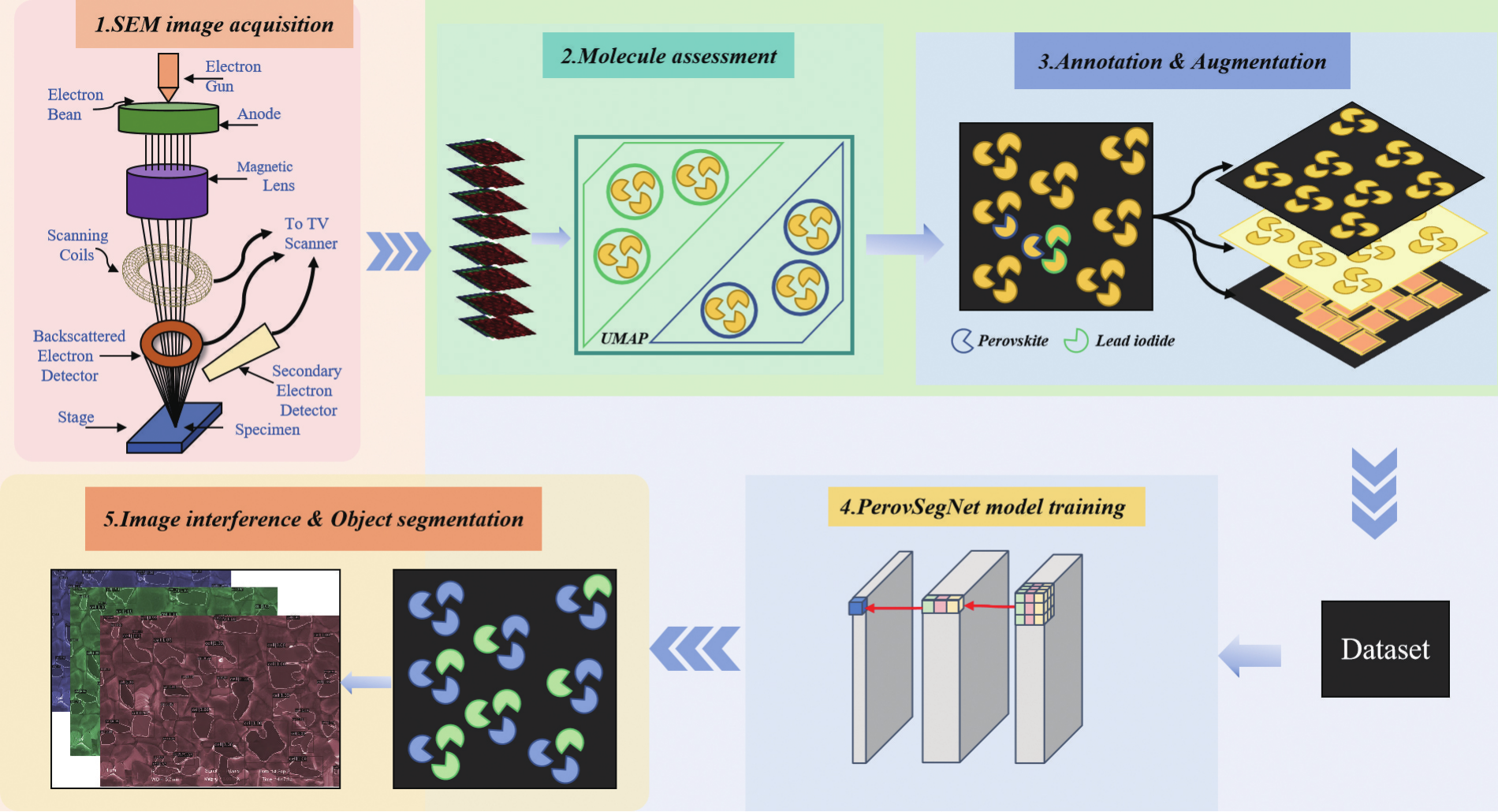}
\caption{Overview of the PerovSegNet-based segmentation workflow. The pipeline begins with SEM image acquisition, followed by the evaluation of structural separability across molecular species using UMAP. A ROI is selected and annotated to construct a training dataset, which is then augmented and used to train PerovSegNet. The trained model performs automated segmentation and predicts pixel-wise classifications for perovskite, lead iodide and defect.}
\label{fig:1}
\end{figure*}

We present PerovSegNet, a high-precision DL model designed for automated segmentation of SEM images of perovskite and lead iodide films. As illustrated in Figure~\ref{fig:1}, the workflow consists of five interconnected steps: (1) SEM image acquisition, (2) evaluation of molecular separability to ensure reliable structural differentiation, (3) selection and annotation of regions of interest (ROI), (4) dataset augmentation and training of PerovSegNet, and (5) automated segmentation and prediction. In the following subsections, we detail the procedures for dataset construction and architectural design of the proposed segmentation model.

\begin{figure*}[htbp]
\centering
\includegraphics[width=1.0\textwidth]{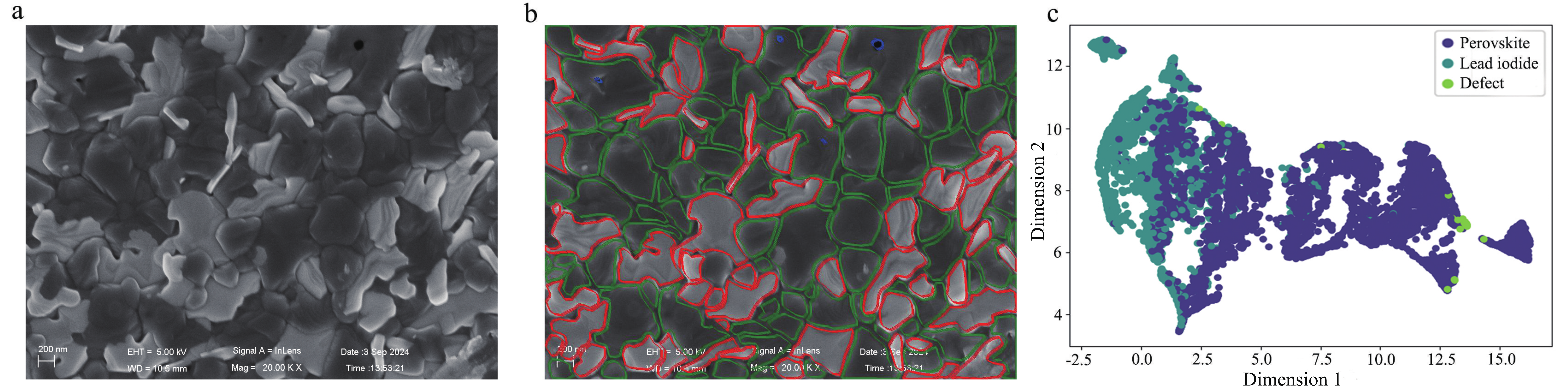}
\caption{Dataset construction and separability evaluation. a) A representative SEM image from PerovData dataset, captured at 20,000 $\times$ magnification, reveals the granular morphology of perovskite films. b) Annotated segmentation map showing individual regions of perovskite (green), lead iodide (red), and defects (blue). c) UMAP projection of the dataset highlighting molecular separability, where clusters correspond to perovskite (purple), lead iodide (dark green), and defects (light green).}
\label{fig:2}
\end{figure*}

\subsection{Dataset Production}
PSCs are typically fabricated via solution-processing techniques such as spin-coating, dip-coating, or spray-coating, followed by thermal annealing to promote perovskite phase formation and crystallization. After film formation, SEM is commonly employed to capture high-resolution images that reveal grain morphology, size distribution, and surface topography at the micro- and nanoscale. A critical challenge in developing automated analysis methods lies in the scarcity of high-quality annotated perovskite SEM datasets. To address this, we constructed the PerovData dataset~\citep{nesteruk2024image}, which comprises 10,994 annotated SEM images of perovskite and lead iodide films, expanded from an initial set of 40 images through the application of several image augmentation techniques~\citep{dash2024addressing}, including random rotations, flips, scaling, and color adjustments. Image annotation was performed using the open-source tool LabelMe\footnote{\url{https://github.com/wkentaro/labelme}}~\citep{russell2008labelme}, generating JSON files containing detailed segmentation information, including object class, pixel-level segmentation masks, and the corresponding location of each segmented region. Raw SEM images are presented in Figure~\ref{fig:2}a, and corresponding annotated masks are shown in Figure~\ref{fig:2}b. The dataset is publicly available at: \href{https://github.com/wlyyj/PerovSegNet-Dataset}{https://github.com/wlyyj/PerovSegNet-Dataset}. During model development, the original dataset, consisting of 40 images, was first randomly partitioned into training, validation, and test sets with an 8:1:1 ratio prior to augmentation to ensure robust performance assessment.

To evaluate the dataset's quality and the separability of perovskite, lead iodide, and defect regions, we employed UMAP, a nonlinear dimensionality reduction technique that preserves both local and global structure in high-dimensional data. By projecting the feature space into two dimensions, UMAP~\citep{mcinnes2018umap} enables intuitive visualization and assessment of class separability. As shown in Figure~\ref{fig:2}c, distinct clusters emerge for the three classes: perovskite, lead iodide, and defects, demonstrating high intra-class cohesion and clear inter-class separation. This supports the reliability of the annotated labels and justifies their suitability for supervised DL segmentation.

\begin{figure}[htbp]
\centering
\includegraphics[width=1.0\linewidth]{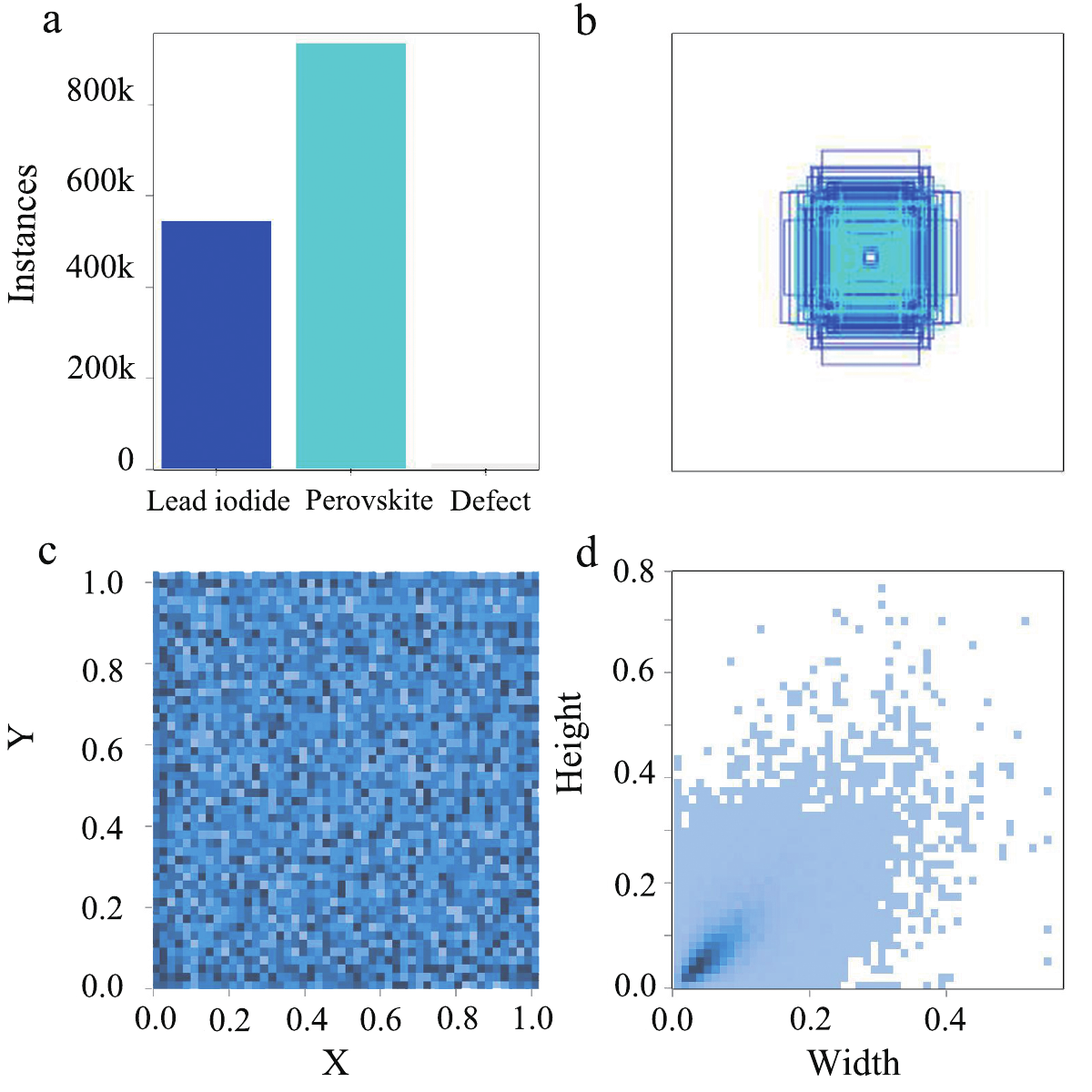}
\caption{Statistical analysis of annotation distribution.  
a) Instance frequency across categories: lead iodide, perovskite, and defect.
b) Heatmap showing the size and number of instances across SEM image regions. 
c) Scatter plot of the position distribution of the center point of instances relative to the entire image.
d) Joint distribution of bounding box width and height.
}
\label{fig:3}
\end{figure}

We further analyzed the geometric and spatial characteristics of annotated instances. Figure~\ref{fig:3}a shows a histogram of instance counts per class: deep blue denotes lead iodide, while light blue denotes perovskite. Figure~\ref{fig:3}b illustrates the width-height distribution of bounding boxes, normalized by aligning their centers, where darker regions indicate higher overlap. This reflects substantial variability in object size. Figure~\ref{fig:3}c shows the spatial distribution of bounding-box centers across the image plane, with axes denoting horizontal and vertical coordinates. Darker regions denote high object density, whereas lighter regions indicate sparsity, revealing a well-balanced spatial distribution. Figure~\ref{fig:3}d shows bounding-box size distribution, with most instances concentrated in smaller ranges, indicating dominance of small objects in the dataset. These statistics provide guidance for training strategies and potential post-processing.

\subsection{Model \& Architecture Enhancement}
The YOLO series of algorithms represents a class of real-time object detection frameworks renowned for their high inference speed and strong detection performance. Among them, YOLOv8 introduces a refined architecture with the C2f module as its backbone, enabling improved feature extraction. However, this default backbone exhibits limitations when applied to high-resolution SEM images, particularly in capturing fine-grained morphological details and managing complex background noise frequently observed in perovskite and lead iodide films. To address these challenges, we propose PerovSegNet, a customized segmentation architecture built upon YOLOv8x-seg and enhanced with two key modules: the ASDCB and the SAD. The ASDCB module replaces the original C2f component, introducing multiple convolutional branches with adaptive receptive fields to enhance feature fusion. This design significantly enhances the network’s ability to discriminate grain boundaries, small particles, and defect regions. In parallel, the SAD module substitutes the traditional pooling-based downsampling operations, which utilizes a dual-path architecture that integrates depthwise separable convolutions with adaptive pooling mechanisms. This approach effectively mitigates common issues in traditional downsampling methods, such as aliasing and limited sensitivity to small-scale textures.

\begin{figure}[htbp]
    \centering
    \includegraphics[width=1.0\linewidth]{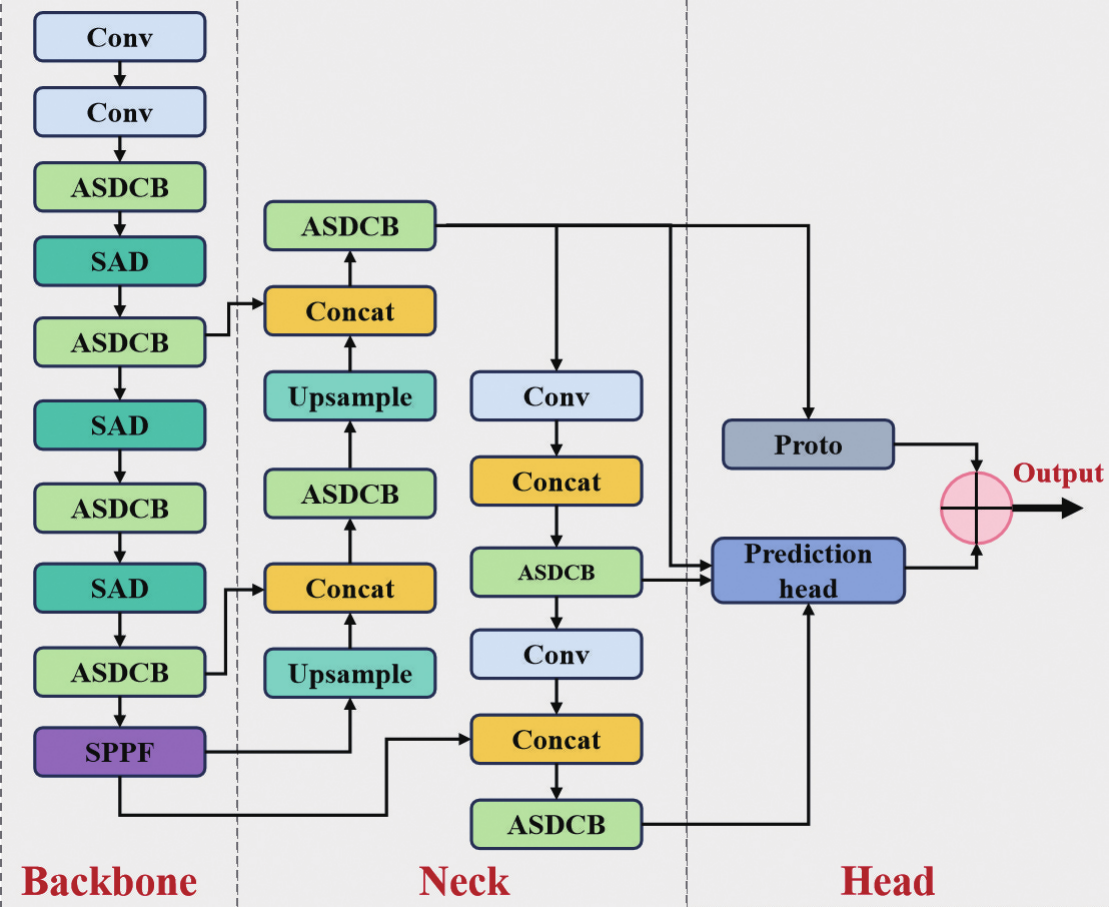}
    \caption{Overall architecture of PerovSegNet. The network consists of three stages: Backbone for hierarchical feature extraction, Neck for multi-scale feature fusion, and Head for segmentation map generation. PerovSegNet tailored for SEM image segmentation of perovskite and lead iodide, extends the YOLOv8x-seg framework by incorporating two customized modules: ASDCB and SAD, which enhance sensitivity to fine-grained morphology and robustness against complex backgrounds.}
    \label{fig:4}
\end{figure}

The overall architecture of PerovSegNet is depicted in Figure~\ref{fig:4}. It follows a typical detection and segmentation pipeline comprising three major components: Backbone, Neck, and Head. The ASDCB-enhanced Backbone is responsible for hierarchical feature extraction; the Neck performs multi-scale feature fusion to integrate semantic and spatial information; and the Head produces high-resolution segmentation masks with improved delineation of perovskite and lead iodide regions. Further details on the structure and functionality of ASDCB and SAD are presented in Sections~\ref{sec:asdcb} and~\ref{sec:SAD}, respectively.

\subsubsection{PerovSegNet}
In the Backbone, we embed the SAD module at each downsampling position, replacing conventional pooling operations. The SAD module employs a dual-path strategy combining depthwise separable convolutions and adaptive pooling mechanisms. The adaptive design allows the module to dynamically capture fine-scale textures and large-scale structures, which are crucial for retaining the integrity of global features and particle boundaries in SEM imagery. Alongside SAD, the ASDCB module is also repeatedly integrated into the Backbone after each convolutional block. ASDCB fuses dilated convolutions with Squeeze-and-Excitation (SE) operations to capture multi-scale contextual features while enhancing local morphological sensitivity. This dual-focus on fine and contextual information improves the model's ability to discriminate between small particles, grain boundaries, and extended defect regions in heterogeneous materials.

In the Neck, the ASDCB module continues to play a critical role in multi-scale feature aggregation. Unlike traditional designs that apply fixed convolutional blocks in the Neck, our incorporation of multi-branch convolutions enables dynamic receptive field adjustment, enhancing the Neck’s ability to propagate high-resolution cues from the backbone while integrating deeper semantic content. Following each feature concatenation and upsampling operation, ASDCB is applied to enrich semantic information and maintain spatial consistency across different resolution levels. This design ensures that both shallow and deep features are adaptively refined before fusion, mitigating the loss of structural detail typically associated with naive upsampling.

In the Head stage, the fused features are passed through the prediction layers, which include the Spatial Pyramid Feature Fusion (SPFF) module and the final segmentation head. SPFF captures hierarchical spatial relationships and improves scale-aware detection. The final 1$\times$1 convolutional layer outputs segmentation maps with class labels, bounding box coordinates, and confidence scores. This architecture enables precise localization and categorization of perovskite grains, lead iodide phases, and defects across a wide range of sizes and shapes.

\subsubsection{Adaptive Shuffle Dilated Convolution Block}
\label{sec:asdcb}
\begin{figure}[htbp]
\centering
\includegraphics[width=0.95\linewidth]{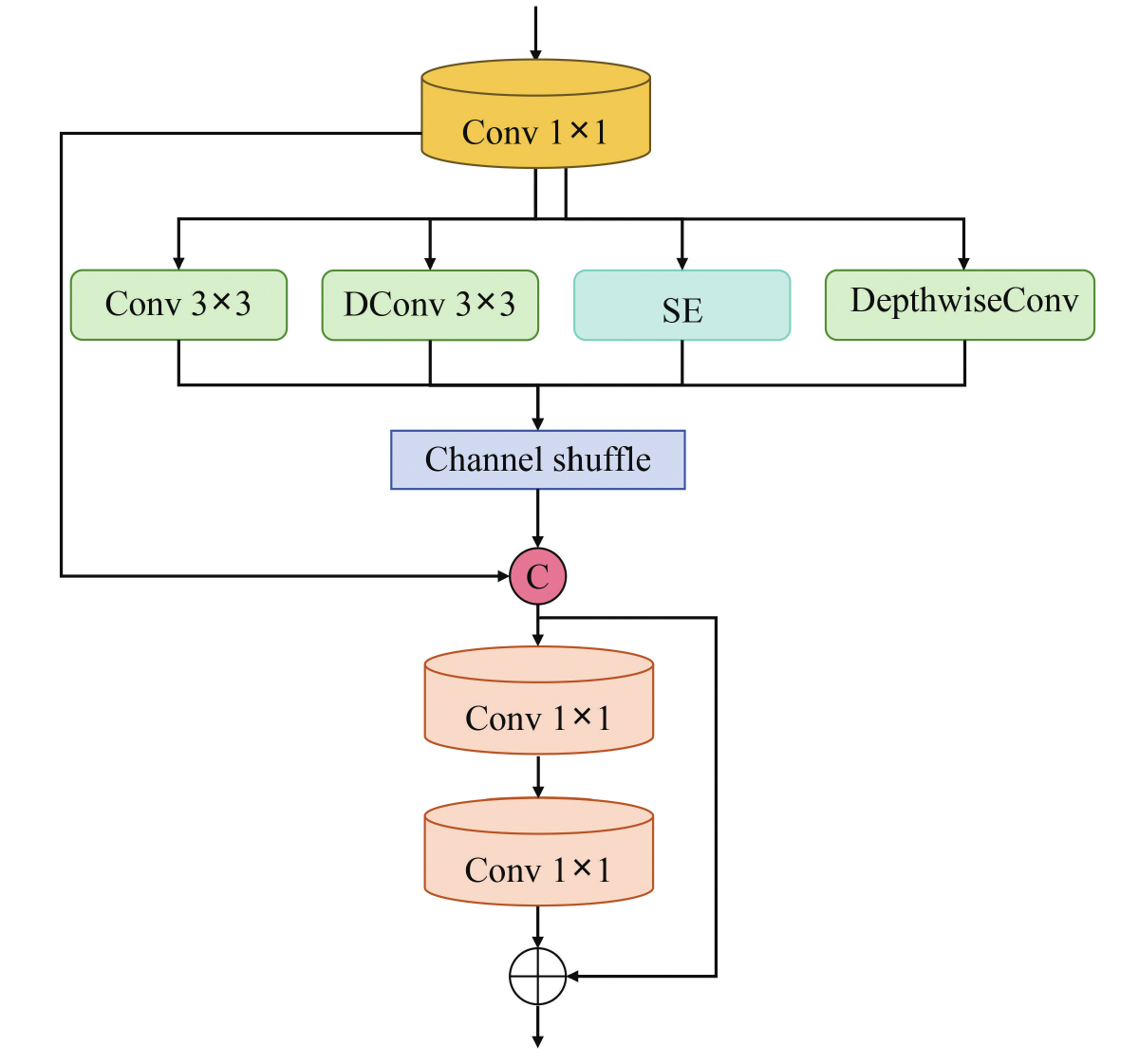}
\caption{Architecture of the ASDCB module. The module is composed of four parallel branches: Standard Convolution, Dilated Convolution, SE layer, and Depthwise Separable Convolution. Each branch is specifically designed to extract complementary features across different scales. The outputs from these branches are fused through a Channel Shuffle operation, which is followed by a ConvFFN for feature integration and transformation. This design enables ASDCB module to enhance the feature representation for complex SEM images of perovskite and lead iodide materials. 
}
\label{fig:5}
\end{figure}

The architecture of the proposed ASDCB is illustrated in Figure~\ref{fig:5}. ASDCB integrates SE attention with dilated and depthwise separable convolutions, and further employs group convolution and channel shuffle strategies to enhance both multi-scale feature aggregation and computational efficiency. Given an input feature map $X$, ASDCB processes it through four parallel branches to extract multi-scale and channel-adaptive features.

The first branch applies standard convolution to extract fundamental spatial features:
\begin{equation}
X_1 = \text{Conv}(X)
\end{equation} 

The second branch employs dilated convolution to enlarge the receptive field and capture broader contextual information:
\begin{equation}
X_2 = \text{DilatedConv}(X, d)
\end{equation} where $d$ represents the dilation rate, allowing the model to capture larger receptive field without increasing parameter count.

The third branch introduces a channel-wise attention mechanism using the SE block to adaptively adjust the weights of each channel, enhancing the network's selectivity for important features. This is implemented via global average pooling (GAP), followed by two fully connected layers:
\vspace{-5pt}
\begin{equation}
S = \sigma(\mathbf{W}_2 \cdot \text{ReLU}(\mathbf{W}_1 \cdot \text{GAP}(X)))
\end{equation}
\vspace{-15pt}
\begin{equation}
X_3 = S \cdot X
\end{equation}
Here, \( \sigma \) is the sigmoid activation, and the parameters \( \mathbf{W}_1 \) and \( \mathbf{W}_2 \) are the weights of the two fully connected layers. \( S \) serves as the learned channel-wise attention vector.

The outputs from all four branches are aggregated:
\begin{equation}
X_{\text{sum}} = X_1 + X_2 + X_3 + X_4
\end{equation}
A channel shuffle operation is then applied to improve cross-branch feature interaction:
\begin{equation}
X' = \text{channel\_shuffle}(X_{\text{sum}}, g)
\end{equation}
where \( g \) denotes the number of groups.
The fused features \( X' \) are passed through a Convolutional Feedforward Network (ConvFFN), which consists of two pointwise convolutions and DropPath regularization, with DropPath improving generalization by randomly omitting entire convolutional paths during training:
\begin{equation}
X_{\text{out}} = \text{ConvFFN}(X')
\end{equation}

\paragraph{Design Rationale and Benefits.}  
The ASDCB module is specifically tailored to address challenges in SEM image segmentation of perovskite and lead iodide films, which often exhibit fine-grained structures, multi-scale morphologies, and complex backgrounds. Conventional convolutional blocks face two major limitations in this context: limited receptive field size and high computational redundancy. ASDCB mitigates these challenges through the following innovations:
\begin{itemize}[noitemsep, leftmargin=8pt]
\item \textbf{Multi-branch convolution for multi-scale feature extraction:} The use of standard, dilated, depthwise, and attention-enhanced convolutions allows ASDCB to capture both localized and contextual features critical for distinguishing small grains and large structural regions.
\item \textbf{Channel shuffle for enhanced feature fusion:} By redistributing channels across branches, the model gains stronger interactions between heterogeneous features, resulting in more expressive representations.
\item \textbf{Computational efficiency via lightweight design:} Depthwise separable convolutions and DropPath reduce computational burden, enabling efficient segmentation on high-resolution SEM data under limited resource constraints.
\end{itemize}

\subsubsection{Separable Adaptive Downsampling Module}\label{sec:SAD}
\begin{figure}[htbp]
\centering
\includegraphics[width=0.95\linewidth]{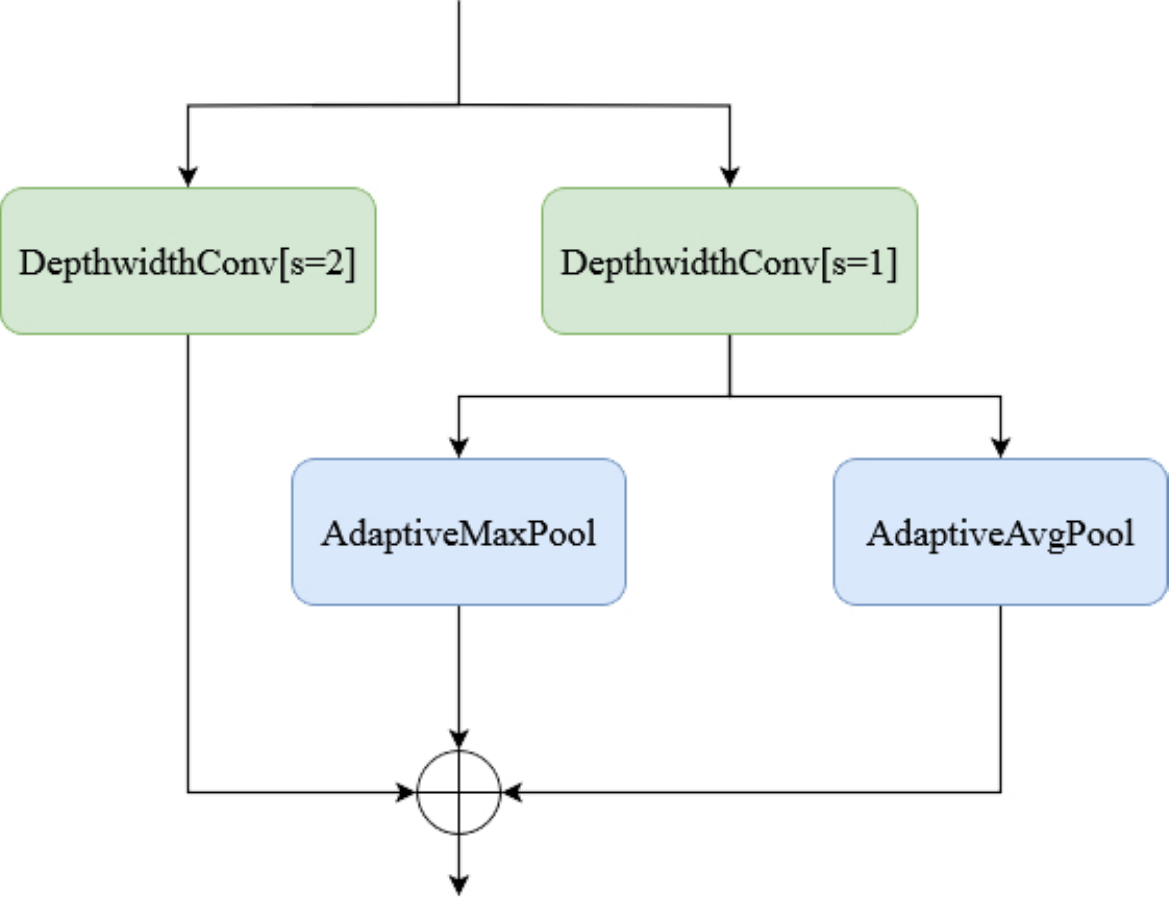}
\caption{
Architecture of SAD module. The module consists of two parallel branches using depthwise separable convolutions with different strides. One branch captures global morphology, while the other is processed through adaptive max pooling and average pooling to preserve high-resolution details. The outputs are concatenated to form a unified feature representation, enabling effective integration of fine details and large-scale structures in SEM image.}
\label{fig:6}
\end{figure}

The architecture of the SAD module is shown in Figure~\ref{fig:6}. It adopts a dual path design, combined with depthwise separable convolution and adaptive pooling, to jointly capture fine scale (high-frequency) textures and large-scale (low-frequency) structures, where fine-scale details correspond to features such as grain boundaries, edges, and textures, while large-scale structures represent wider morphology and overall film uniformity~\citep{li2020frequency}. Depthwise separable convolutions process each channel independently, making them particularly effective at preserving subtle intensity differences between neighboring pixels, for example, those observed at crystal boundaries or particle edges, thus retaining critical microstructural information. On this basis, the left branch uses stride $s=2$ for downsampling, focusing on capturing global features or coarse-grained structural information, such as large-scale shape and overall uniformity. The right branch uses stride $s=1$ to maintain the spatial resolution of the input feature map, and enhances local feature extraction capability through adaptive max pooling and adaptive average pooling. Adaptive max pooling highlights significant local structural features, while adaptive average pooling helps aggregate smoother contextual information. Together, these two pathways enable the SAD module to integrate information across different spatial scales, thereby improving segmentation accuracy for complex perovskite and lead iodide morphologies.

Given an input feature map \( X \in \mathbb{R}^{C \times H \times W} \), where \( C \) is the number of channels and \( H \), \( W \) are the spatial dimensions, the SAD module first applies two depthwise separable convolutions with different strides: one with stride $ s=2 $ for downsampling, and another with stride $ s=1 $ to retain high-resolution details. These convolutions can be expressed as:
\begin{equation}
X_{\text{1}} = \text{DepthwidthConv}_1(X) \quad \text{with} \quad s = 2
\end{equation}
\vspace{-15pt}
\begin{equation}
X_{\text{2}} = \text{DepthwidthConv}_2(X) \quad \text{with} \quad s = 1
\end{equation}

Subsequently, the output feature map is further processed by adaptive max pooling and adaptive average pooling, which adjust pooling regions based on the input to capture both fine and coarse features:
\begin{equation}
X_{\text{max}} = \text{AdaptiveMaxPool}(X_{\text{2}})
\end{equation}
\vspace{-15pt}
\begin{equation}
X_{\text{avg}} = \text{AdaptiveAvgPool}(X_{\text{2}})
\end{equation}

The results from both pooling operations are then concatenated to form a unified feature map:
\begin{equation}
X_{\text{concatenated}} = \text{Concat}(X_{\text{1}}, X_{\text{max}}, X_{\text{avg}})
\end{equation}

The key innovation of the SAD module lies in combining depthwise separable convolutions with adaptive pooling for downsampling, instead of relying solely on static pooling. Traditional pooling methods are computationally efficient but often discard fine details and boundary precision, which are essential in SEM image segmentation. By contrast, the SAD module balances local and global representations, retaining subtle boundaries while preserving large-scale morphology. This adaptive strategy is particularly effective for capturing the intricate features of perovskite and lead iodide microstructures, thereby enhancing segmentation accuracy.

\section{Experiments}
\subsection{Network Training}
All experiments were conducted on a CentOS Linux 7.6.1810 (Core) system, with the software and hardware configurations summarized in Table~\ref{tab:1}. Model development and training were carried out using Visual Studio Code 1.94, PyTorch 2.2.2, and CUDA 12.0. The computing platform was equipped with an NVIDIA A100 GPU (80 GB graphics memory) and 1.0 TB of system memory.

\begin{table}[b]
\Large %
\centering
\caption{Hardware and software setup used for training.}
\label{tab:1}
\resizebox{0.95\linewidth}{!}{
\renewcommand{\arraystretch}{1.0}
\fontfamily{ptm}\selectfont
\begin{tabular}{>{\centering\arraybackslash}p{5cm} >{\centering\arraybackslash}p{7cm}} 
\hline
Name & Configuration/Version \\
\hline
Operating system & CentOS Linux 7.6.1810(Core) \\
CPU & AMD EPYC 7543 (32-core) \\
GPU & NVIDIA A100 80GB PCIe \\
Memory & 1.0TB \\
Graphics memory & 80 GB \\
IDE & Visual Studio Code 1.94 \\
Framework & PyTorch 2.2.2 \\
CUDA & CUDA 12.0 \\
cuDNN & cuDNN 8.9.0 \\
Python Version & Python 3.8.19 \\
\hline
\end{tabular}
}
\end{table}
\begin{table}[b]
\Large %
\centering
\caption{Hyperparameter settings for network training.}
\label{tab:2}
\resizebox{1.0\linewidth}{!}{
\renewcommand{\arraystretch}{1.0}
\fontfamily{ptm}\selectfont
\begin{tabular}{>{\centering\arraybackslash}p{5cm} >{\centering\arraybackslash}p{7cm}} 
\hline
Parameter & Value \\ 
\hline
Initial learning rate & 0.01 \\ 
Epochs & 300 \\ 
Batch size & 4 \\ 
Momentum & 0.937 \\ 
Weight decay & 0.0005 \\ 
Warm-up epochs & 3.0 \\ 
Warm-up momentum & 0.8 \\ 
Warm-up bias learning rate & 0.1 \\ 
Input image size & $640 \times 640$ \\ 
Number of classes & 3 \\ 
Optimizer & AdamW \\ 
\hline
\end{tabular}
}
\end{table}

In this study, the SEM images of perovskite and lead iodide were resized to $640\times640$ pixels to meet the requirements of the YOLOv8 model. Due to YOLOv8's use of adaptive scaling and padding techniques at the input stage, the actual preprocessing size of the input images may not strictly be $640\times640$, but rather have one side resized to 640, with the other side determined by padding according to the aspect ratio of the original image.

Training was conducted for 300 epochs with a batch size of 4, balancing convergence and resource constraints. The initial learning rate was set to 0.01 with a weight decay of 0.0005. Optimization was performed using the AdamW optimizer. Warm-up strategies, including momentum warm-up and learning rate bias warm-up, were employed during the initial training phase to stabilize convergence. The detailed hyperparameter settings are shown in Table~\ref{tab:2}.

\subsection{Comparison with Other Models}
\begin{table*}[b]
\centering
\caption{Comparison of different models on the PerovData dataset. The best results are highlighted in \textbf{bold}.}
\resizebox{0.85\linewidth}{!}{
\fontfamily{ptm}\selectfont
\begin{tabular}{cccccc} 
\hline
Model & mAP@0.5 & mAP@0.5:0.95 & Params (M) & GFLOPs (G)  \\ 
\hline
Cascade Mask R-CNN ($\times$101)    & 80.4  & 77.5  & 122  & 1493  \\ 
Mask R-CNN (Swin-T)   & 82.5  & 77.9  & 48   & 267   \\ 
Mask R-CNN ($\times$101)     & 80.6  & 78.0  & 44   & 447   \\ 
Cascade InternImage-XL & 81.9  & 79.8  & 387  & 1782  \\ 
\textbf{PerovSegNet}           & \textbf{87.3}  & \textbf{80.0}  & \textbf{79}   & \textbf{265}   \\ 
\hline
\end{tabular}
}
\label{tab:3}
\end{table*}
\begin{table*}[b]
\centering
\caption{Ablation study of PerovSegNet on the PerovData dataset. The best results are highlighted in \textbf{bold}.}
\resizebox{1.0\textwidth}{!}{
\fontfamily{ptm}\selectfont
\begin{tabular}{ccccccccc} 
\hline
Number & SAD & ASDCB & Precision & Recall & mAP@0.5  & mAP@0.5:0.95 & Params(M) & GFLOPs(G)  \\ 
\hline
1  & --- & ---   & 97.30 & 75.31 & 83.17 & 76.48 & 104.67  & 354.9  \\ 
2  & \checkmark & --- & 97.38 & 76.99 & 86.30 & 79.31 & 96.49  & 340.1  \\ 
3   & --- & \checkmark & 97.94 & 78.04 & 86.42 & 78.23 & 87.28 & 280.1    \\ 
4   & \checkmark & \checkmark & \textbf{98.11} & 
\textbf{76.30} & \textbf{87.25} & \textbf{79.96} & \textbf{79.10}  & \textbf{265.4}   \\ 
\hline
\end{tabular}
}
\label{tab:4}
\end{table*}

\subsection{Evaluation Metrics}
The performance of the proposed PerovSegNet model is quantitatively assessed using six commonly used metrics: Precision, Recall, mean Average Precision at IoU=0.5 (mAP@0.5)~\citep{everingham2010pascal}, mean Average Precision at IoU thresholds from 0.5 to 0.95 (mAP@0.5:0.95), number of trainable parameters, and computational complexity in Giga Floating Point Operations (GFLOPs). These metrics collectively evaluate the detection accuracy, segmentation quality, and computational efficiency of the model. Precision quantifies the proportion of correctly predicted positive instances out of all predicted positive instances, while Recall calculates the proportion of true positive instances out of all actual positive instances, offering insights into the model's ability to identify and classify relevant objects. The mAP@0.5 metric measures the average precision across all classes at an IoU threshold of 0.5, providing a comprehensive assessment of detection performance. The mAP@0.5:0.95 metric extends this by evaluating the average precision across IoU thresholds from 0.5 to 0.95, offering a more stringent measure of detection quality that takes into account the model’s ability to match predicted and ground truth objects at varying degrees of overlap. The number of Parameters reflects the overall model complexity, and GFLOPs indicates the computational cost during inference. These metrics enable a thorough evaluation of the model's segmentation capabilities on the SEM dataset of perovskite and lead iodide films, considering both detection fidelity and computational feasibility.

\section{Results and Analyze}

To evaluate the effectiveness of the proposed PerovSegNet, we conducted comprehensive comparisons against several widely adopted segmentation models, including Mask R-CNN, Cascade Mask R-CNN, and the state-of-the-art Cascade InternImage-XL. All models were trained and tested on the PerovData dataset, which contains SEM images of perovskite and lead iodide materials. The experiments were conducted using the MMDet\footnote{\url{https://github.com/open-mmlab/mmdetection}} training framework, which provides an efficient and flexible environment for implementing and evaluating object detection and segmentation models. For all models, we used the Adam optimizer with an initial learning rate of 0.001, decayed using a stepLR scheduler with a decay factor of 0.1 every 10 epochs. Each model was trained for a total of 120 epochs with a batch size of 16. We employed a weight decay of 0.0001 and gradient clipping with a threshold of 5 to prevent exploding gradients. Table~\ref{tab:3} presents quantitative results in terms of mAP@0.5, mAP@0.5:0.95, parameter count, and GFLOPs. PerovSegNet achieved an mAP@0.5 of 86.8\%, substantially outperforming all baselines. Furthermore, it demonstrated superior computational efficiency with only 87 million parameters and 280 GFLOPs. Compared to models such as Cascade Mask R-CNN and Cascade InternImage-XL, which require significantly more resources, PerovSegNet striked a favorable balance between segmentation accuracy and model complexity.

\subsection{Ablation Analysis}

To assess the individual contributions of the proposed SAD and ASDCB modules, we performed ablation experiments based on the YOLOv8x-seg backbone. Table~\ref{tab:4} summarizes the results in terms of Precision, Recall, mAP@0.5, mAP@0.5:0.95, parameter count, and GFLOPs. The baseline YOLOv8x-seg achieved an mAP@0.5 of 83.17\%. Adding the SAD module improved performance to 86.30\%, highlighting the benefits of its dual-path design in preserving fine-scale details. Incorporating ASDCB alone further boosts mAP@0.5 to 86.42\%, demonstrating its ability to capture complex multi-scale features. When both modules were combined, the model achieved the highest mAP@0.5 of 87.25\% and mAP@0.5:0.95 of 79.96\%, while reducing the parameter count and GFLOPs by 24\% and 25\%, respectively, compared to the baseline.

\begin{figure*}[htbp]
    \centering
    \includegraphics[width=1.0\linewidth]{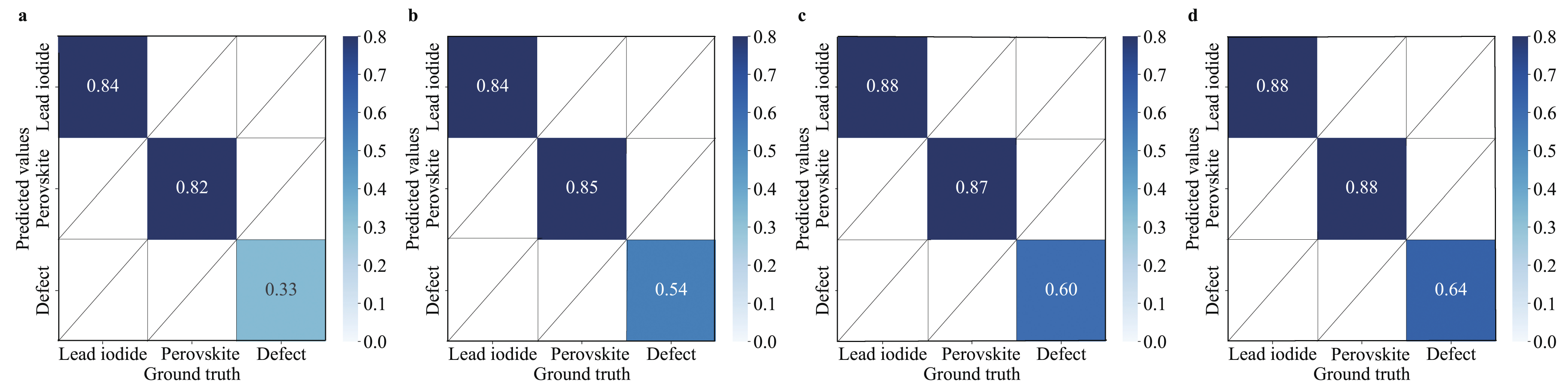}
    \caption{Normalized confusion matrices from ablation study. a) YOLOv8x-seg baseline, b) YOLOv8x-seg + ASDCB, c) YOLOv8x-seg + SAD, and d) YOLOv8x-seg + ASDCB + SAD (PerovSegNet). The results highlight improved class-wise segmentation accuracy, especially for defect regions.}
    \label{fig:7}
\end{figure*}
We further analyzed class-specific segmentation accuracy for lead iodide, perovskite and defect regions, and the results are shown in Figure~\ref{fig:7}. The segmentation accuracy of lead iodide accuracy improved from 84\% (baseline) to 88\% with both modules, while the segmentation accuracy of perovskite accuracy increased from 82\% to 88\%. The largest gain was observed for defect segmentation, where accuracy rose from 33\% in the baseline model to 64\% with both modules. These results confirm that ASDCB strengthens multi-scale feature extraction, whereas SAD preserves fine-scale details critical for defect regions. Their combination enables robust segmentation of both crystalline domains and small irregular features, substantially enhancing performance in SEM image analysis.

\subsection{Qualitative Evaluation: Segmentation of Perovskite and Lead Iodide}
\begin{figure}[htbp]
    \centering
    \includegraphics[width=0.95\linewidth]{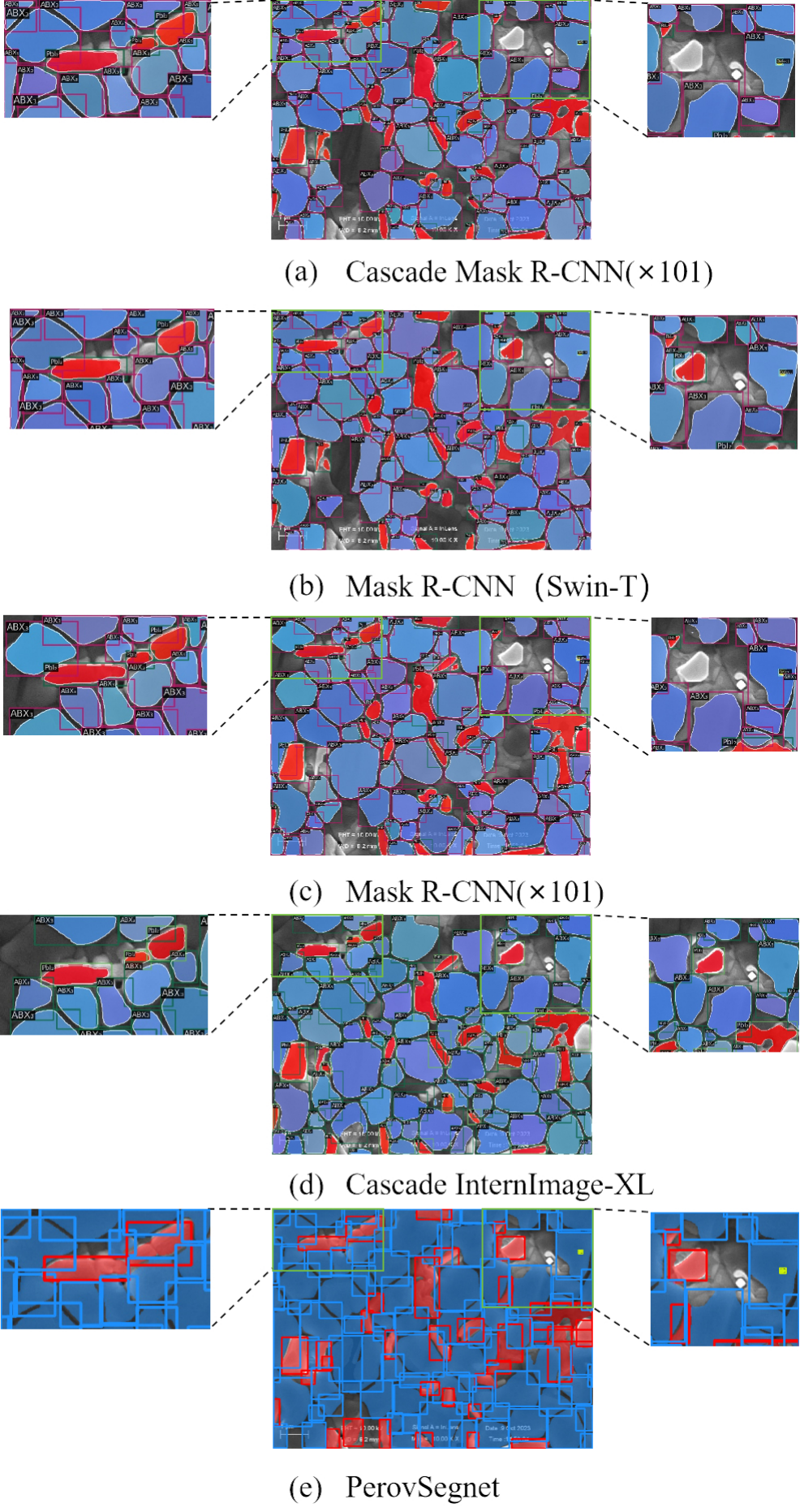}
    \caption{Qualitative comparison of segmentation on representative SEM image. (a) Cascade Mask R-CNN ($\times$101), (b) Mask R-CNN (Swin-T), (c) Mask R-CNN ($\times$101), (d) Cascade InternImage-XL, and (e) PerovSegNet (ours). Cascade/Mask variants exhibit leakage in lead iodide regions (a,c) and missed small defects (d), while Swin-T shows leakage in larger perovskite areas due to limited global context (b). PerovSegNet achieves cleaner boundaries and fewer missed detections, particularly for small lead iodide clusters and defect regions.}
    \label{fig:8}
\end{figure}

To further validate model behavior beyond aggregate metrics, we performed a qualitative comparison on representative SEM images that include challenging scenarios, densely packed grains, irregular boundaries, and small defect regions (Figure~\ref{fig:8}). In Figure~\ref{fig:8}a, Cascade Mask R-CNN ($\times$101) exhibited segmentation leakage in lead iodide regions, predicted masks spill across grain boundaries, revealing the difficulty of purely convolutional backbones in adhering to fine-scale, irregular contours. Figure~\ref{fig:8}b showed that Mask R-CNN (Swin-T) recovers some fine structures due to its Transformer features. Nevertheless, despite offering longer-range modeling than standard CNNs, its windowed attention under-captures global context, leading to leakage in larger perovskite domains. Figure~\ref{fig:8}c (Mask R-CNN, $\times$101 with convolutional backbones) similarly under-segmented thin lead iodide clusters and yielded high false negatives in low-contrast defect areas. In Figure~\ref{fig:8}d, Cascade InternImage-XL reduced certain boundary artifacts through multi-stage refinement but still misses small, low-contrast defects, indicating that refinement alone cannot reconcile local detail with broader morphological context. By contrast, Figure~\ref{fig:8}e (PerovSegNet) delineated sharper boundaries, separates densely packed grains more reliably, and markedly reduces missed lead iodide clusters and defect regions. These gains are consistent with the ablation and class-wise results and stem from the complementary modules: ASDCB strengthens multi-scale representation and boundary localization for small/irregular objects, whereas SAD preserves fine-scale textures and large-scale structures, thereby suppressing leakage and false negatives under complex SEM morphologies.

\section{Discussion}
\begin{figure}[htbp]
    \centering
    \includegraphics[width=0.95\linewidth]{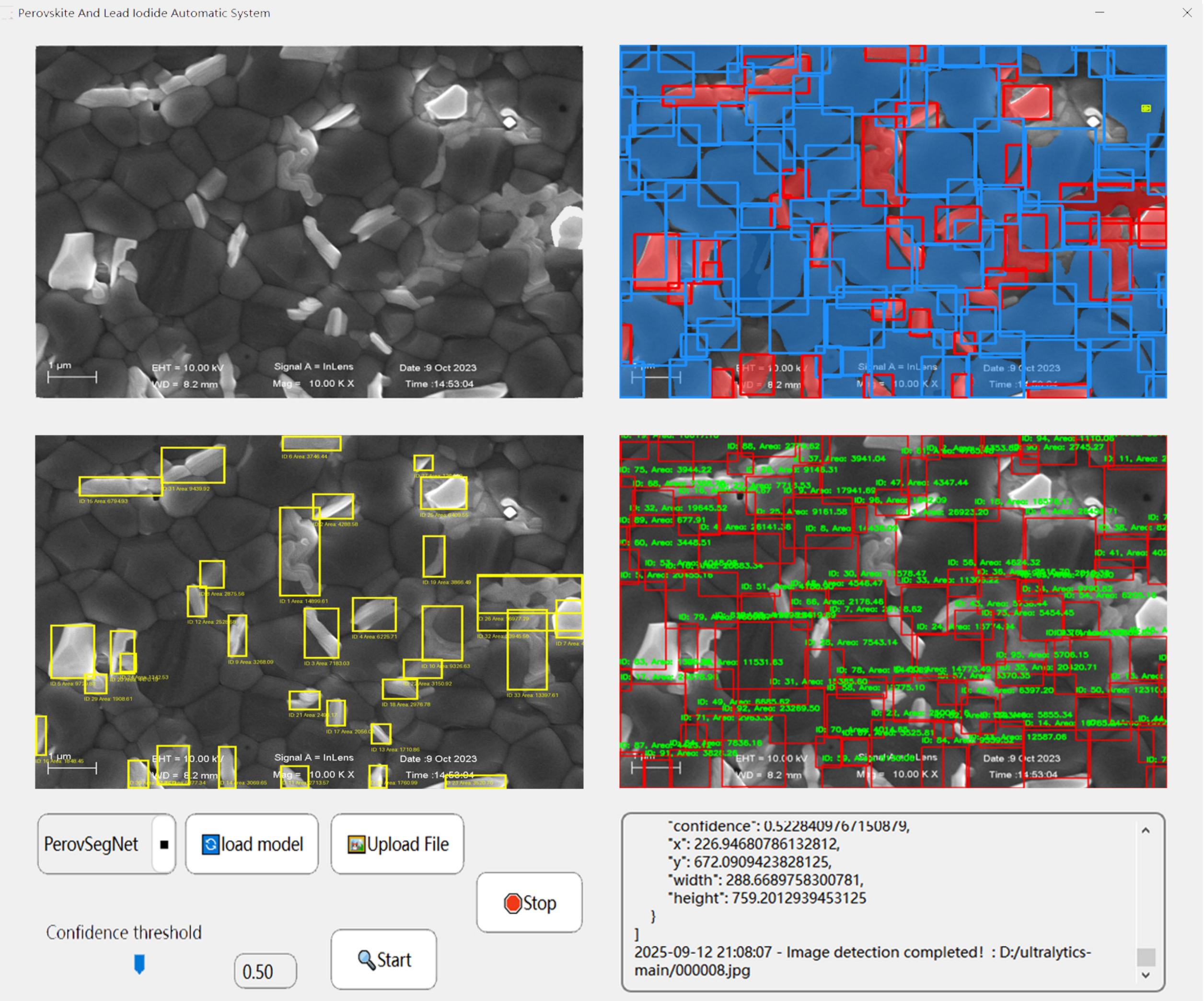}
    \caption{Automated segmentation and quantification platform. Top-left: original SEM image. Top-right: segmentation map (perovskite in blue, lead iodide in red, defects in yellow). 
    Middle-left: automated statistics reporting lead iodide. Middle-right: automated statistics reporting perovskite. 
    Bottom: analysis panel with runtime settings and logs.  
    The outputs provide per-image metrics for process feedback and quality control.}
    \label{fig:9}
\end{figure}

\begin{figure}[htbp]
    \centering
    \includegraphics[width=0.95\linewidth]{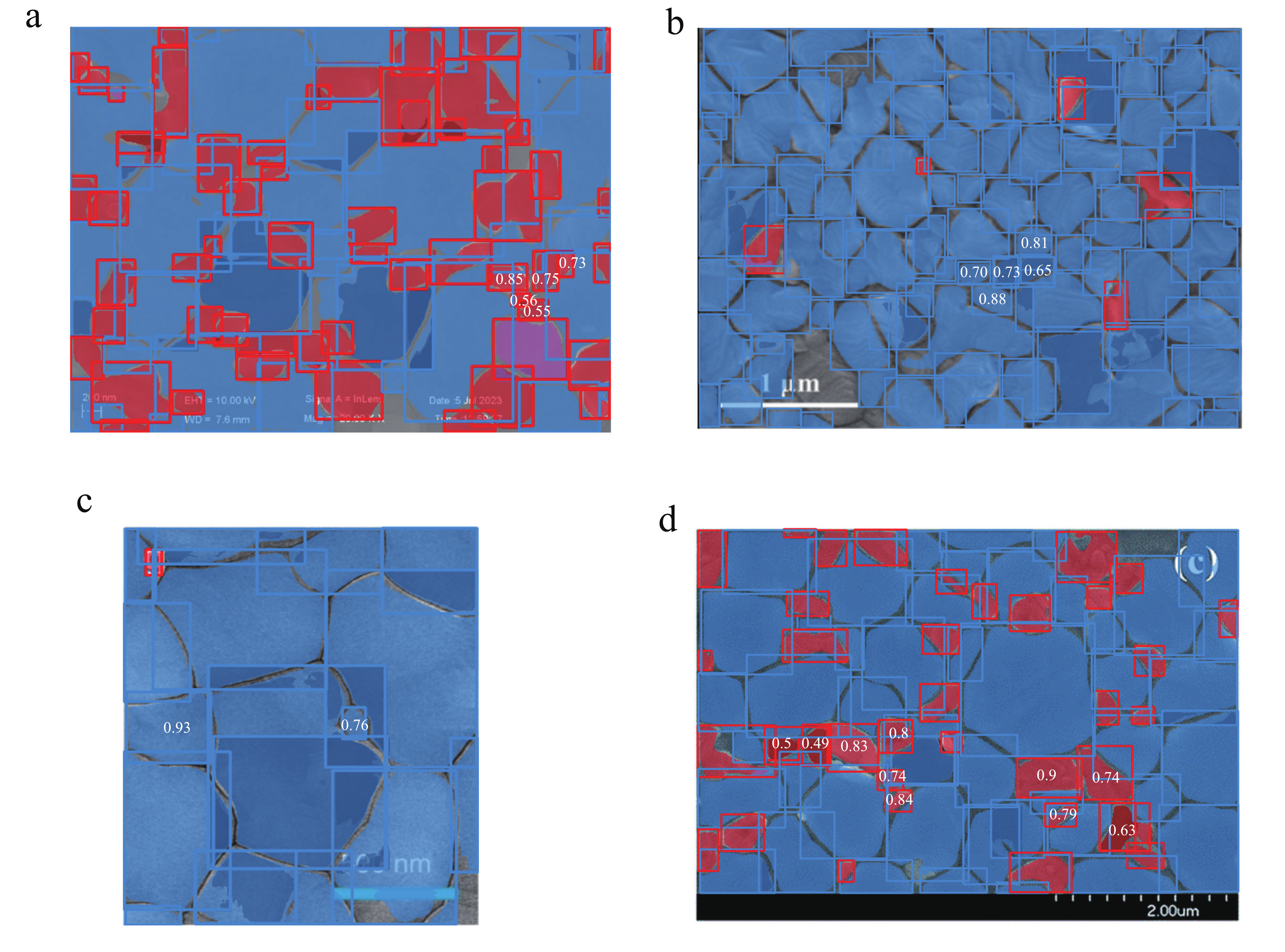}
    \caption{PerovSegNet segmentation results across varying SEM resolutions. SEM images at different magnifications: (a) 200 nm, (b) 1 µm, (c) 500 nm, and (d) 2 µm. The red and blue outlines represent segmented regions of interest, corresponding to lead iodide and  perovskite, while the white font labeled numbers display the confidence level of the segmented area.}
    \label{fig:10}
\end{figure}
PerovSegNet advances SEM image analysis not only by improving segmentation accuracy but also by producing quantitative descriptors that are directly connected to material structure. The framework targets SEM-specific challenges, multi-scale grain boundaries, irregular morphologies, and low-contrast features, through two complementary modules ASDCB and SAD. The quantitative and qualitative gains in the ablation study can translate into more reliable downstream microstructural metrics.

\noindent \textbf{System-level demonstration (Figure~\ref{fig:9}).} Figure~\ref{fig:9} illustrates the end-to-end pipeline in an engineering context. Starting from a raw SEM image, PerovSegNet generates instance masks for perovskite, lead iodide, and defect, together with per-image statistics (areas in $\mu$m$^2$ and counts). In the example shown, perovskite and lead iodide domains are clearly delineated and small, low-contrast defects are retained, consistent with the failure modes observed in our comparisons against representative CNN baselines. The tabulated outputs are readily usable for analysis and enable derivation of additional indicators such as grain-size distributions and perimeter-to-area ratios (shape complexity), which can inform hypothesis-driven process adjustments (e.g., annealing schedule, precursor stoichiometry, antisolvent timing) to steer films toward target morphologies.

%We select approximately 25 papers published between 2022 and 2025 that contain SEM images and corresponding PCE values. We then used our Perovsegnet model to segment the images and extract structural information for correlation analysis.

\noindent \textbf{Scale-invariant SEM analysis with PerovSegNet (Figure~\ref{fig:10}).} To evaluate the resolution capability of our model across different electron microscope scales, we selected SEM images of perovskite at various magnifications from peer-reviewed papers for comparison. As shown in Figure~\ref{fig:10}, we have highlighted regions that are typically challenging for manual annotation. Our model effectively processes SEM images at multiple magnifications, ranging from high-resolution scans (e.g., 500 nm in image c) to lower magnifications (e.g., 1 µm in image b), demonstrating exceptional accuracy (averaging over 75\%) in identifying areas with grain overlap. The model consistently identifies perovskite and lead iodide features with high precision. This performance is primarily attributed to the design of the SAD module, which allows for the simultaneous capture of fine-scale textures and large-scale structures, particularly in regions prone to ambiguity. Additionally, the ASDCB module enhances multi-scale feature fusion, improving accuracy in processing regions of dense SEM data. The resolution-invariance capability of our model is crucial for adapting to SEM data collected under various experimental conditions. It ensures that, regardless of image resolution or scale, the model can reliably generate morphological descriptors for further correlation with photovoltaic efficiency (PCE), facilitating future research.

\begin{figure}[htbp]
    \centering
    \includegraphics[width=0.95\linewidth]{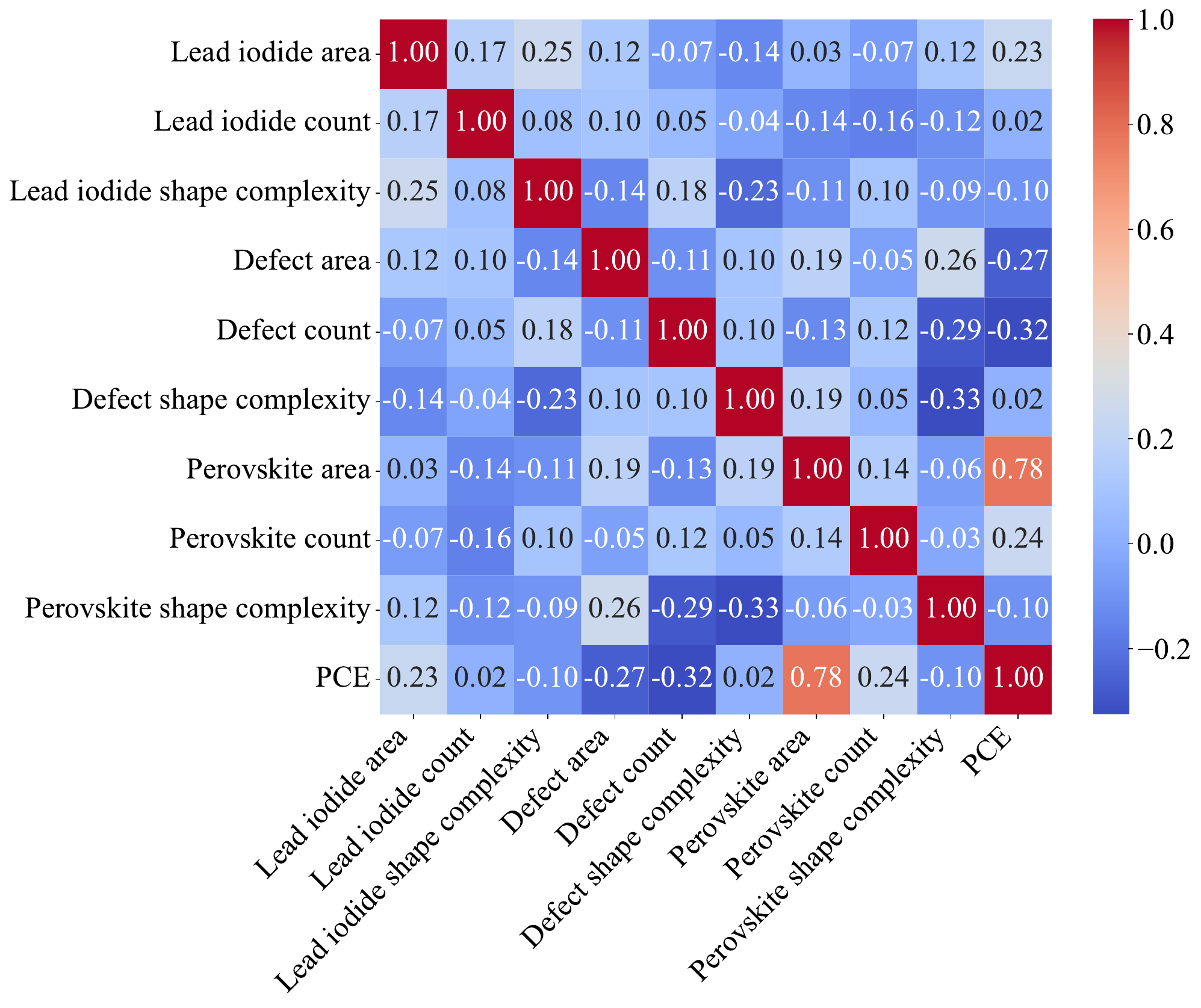}
    \caption{Correlation heatmap between image-derived microstructural descriptors and device PCE. 
    Features include the area, count and shape complexity of lead iodide, perovskite and defect. Cells show Pearson correlation coefficients, where red indicates positive correlation and blue indicates negative correlation.}
    \label{fig:11}
\end{figure}

\noindent \textbf{Linking microstructure to device performance (Figure~\ref{fig:11}).} To demonstrate the feasibility of linking image-derived structural descriptors with photovoltaic performance, we collected a representative set of approximately 25 peer-reviewed papers published between 2022 and 2025 that report both high-resolution SEM images and corresponding PCE values. These studies were selected to ensure the necessary level of detail for accurate structural analysis and the availability of device performance data. We then applied our PerovSegNet model to segment the SEM images and extract quantitative morphological features for correlation analysis. As shown in Figure~\ref{fig:11}, the correlation heatmap (Pearson $r$) highlights the relationships between structural descriptors and PCE. Perovskite area shows a strong positive correlation with PCE ($r$ = 0.78), while perovskite shape complexity (perimeter-to-area ratio) exhibits a slight negative correlation ($r$ = -0.10). Together, these results indicate that both the extent and the regularity of perovskite coverage play critical roles in device performance. Larger perovskite domains enhance light absorption and carrier transport, whereas more uniform and regular grain morphology reduces defect-assisted recombination and facilitates smoother charge pathways. These structural features jointly enable more effective utilization of photo-generated carriers, thereby contributing to higher PCE. Lead iodide area shows a mild positive correlation with PCE ($r$ = 0.23), reflecting the complex and dual role of lead iodide in perovskite films. When present in a controlled and discontinuous form, lead iodide promotes perovskite crystallization and enhances interfacial stability, which improves charge transport across layers. However, excessive or continuous lead iodide residues can hinder carrier extraction and accelerate degradation, leading to efficiency loss. The observed mild positive correlation is thus consistent with beneficial effects of moderate lead iodide content, but underscores the need to carefully balance its presence to maximize PCE. Defect-related metrics show negative correlations with PCE, with defect area and defect count correlating at $r$ = -0.27 and $r$ = -0.32, respectively. Defects such as cracks, pinholes, and irregular grain boundaries act as non-radiative recombination centers, introducing trap states that shorten carrier lifetimes. These structural irregularities disrupt charge continuity, increase recombination losses, and reduce current output. As a result, films with higher defect densities consistently exhibit lower PCE, confirming the importance of accurate defect detection and reduction during fabrication.

\noindent \textbf{From metrics to process guidance.} Together, Figures~\ref{fig:9} and \ref{fig:10} demonstrate how segmentation outputs can be transformed into predictive indicators for performance-oriented process optimization. By linking microstructural descriptors such as perovskite coverage, lead iodide distribution, and defect density to device-level outcomes, PerovSegNet provides a pathway toward data-driven feedback loops: (i) acquire SEM, (ii) segment and quantify, (iii) consult correlations to anticipate performance impact, and (iv) adjust process parameters to steer morphology toward favorable regimes. While this loop remains prospective, it illustrates the potential for automated SEM analysis to transition from descriptive imaging to actionable process control.

\noindent \textbf{Limitations and scope.} The reported correlations are observational and do not imply causality. Other factors such as composition or interface engineering may influence both morphology and PCE. Segmentation errors and class imbalance, particularly for defect classes, can propagate into derived metrics. Imaging conditions (voltage, magnification, calibration) also impact quantitative outputs. Addressing these issues will require larger, more diverse datasets, uncertainty quantification, and cross-magnification validation. Although computational efficiency is improved relative to strong baselines, further acceleration will be necessary for integration into in situ or inline workflows.

\section{Conclusion}
Artificial intelligence-based image processing is becoming indispensable for analyzing and optimizing perovskite films. In this work, we develop a YOLOv8-seg-based segmentation framework PerovSegNet customized for SEM with perovskite and lead iodide. By proposing two tailored modules, ASDCB strengthening multi-scale feature extraction and boundary localization, and SAD preserving both fine-scale textures and large-scale structures, the model achieves state-of-the-art segmentation accuracy while reducing computational cost. Evaluations on the curated PerovData dataset demonstrate that PerovSegNet surpasses strong baselines, such as Mask R-CNN (Swin-T backbone) and Cascade InternImage-XL. Beyond segmentation, PerovSegNet also provides quantitative descriptors, including the area and count of perovskite and lead iodide, that are physically interpretable and directly linked to photovoltaic performance. These capabilities establish a foundation for data-informed process optimization, where microstructural features such as perovskite area, lead iodide distribution, and defect density can be systematically monitored and adjusted to enhance device efficiency and stability.

Despite these advances, challenges remain in extending the framework to other material systems and in further accelerating inference for in situ and inline applications. Looking forward, the lightweight design and robust quantification capabilities of PerovSegNet suggest its potential not only for laboratory-scale research but also for high-throughput screening and real-time quality control in perovskite manufacturing, thereby helping to accelerate the path toward scalable photovoltaic technologies.

\section*{Data availability}
The datasets generated and/or analyzed during the study are available on GitHub (https://github.com/wlyyj/PerovSegNet-Dataset). Source data are provided with this paper.

\section*{Code availability}
The algorithm is written in Python 3.8.0. The corresponding source codes and scripts are available through GitHub at (https://github.com/wlyyj/PerovSegNet/tree/master). 

\section*{Acknowledgment}
This work was supported by the National Natural Science Foundation of China under Grant 62404136. 

\section*{Conflicts of interest}
The authors have no conflicts to disclose. 

\bibliographystyle{plain}  % 选择参考文献的样式，如plain、ieeetr、unsrt等
\bibliography{main}

\end{document}